\documentclass[12pt]{iopart}
\usepackage{iopams}
\usepackage{color} 
\usepackage{graphicx}
\usepackage{epstopdf} 

\begin{document}

\title
{Wave Triad with Forcings as a Nambu System}
\author{Richard Blender}

\address{Universit\"at Hamburg, Meteorologisches Institut, Center for Earth System Research and Sustainability (CEN), Bundesstr. 55, D-20146 Hamburg, Germany}
\ead{richard.blender@uni-hamburg.de}

\author{Joscha Fregin}

\address{Technische Universit\"at Hamburg, Institut f\"ur Mathematik (E-10), Am Schwarzenberg-Campus 3, D-21073 Hamburg, Germany}

\vspace{10pt}

\begin{abstract}
The dynamics of an ideal wave triad with real amplitudes has a well-known Nambu representation with energy and enstrophy as conservation laws. Here we derive Nambu representations for systems with constant forcings. These equations have been applied to triads of Rossby-Haurwitz waves in the atmosphere where they are forced with orography. The conservation laws are based on relations for the unforced amplitudes  and a Hamiltonian given by the total energy plus terms involving the unforced amplitudes.  The forcing of the unstable wavenumber causes a recharge cycle.  

 \vspace{2pc}


\end{abstract}

{\tiny \today}

%
%
%

\section{Introduction}

%
Nambu  \cite{Nambu1973} has suggested an extension of conservative dynamical systems which is based on the Liouville Theorem. In the simplest nontrivial case this pertains to a system with  three degrees of freedom with a second conservation law in addition to a Hamiltonian. The dynamics is given in terms of a Nambu bracket which generalizes Lie-Poisson brackets \cite{Takhtajan1994}. Casimirs in this theory are given by the second conservation law in the Nambu bracket. The concept of Nambu mechanics has been extended to continuous hydrodynamic systems with a finite number of conserved integrals \cite{NevirBlender1993,NevirSommer2009}. 

In geophysical flows the weakly nonlinear interaction of Rossby-Haurwitz wave is considered to be a main constituent of atmospheric turbulence (see e.g. \cite{Kartashova_2008}). Three waves can build  a resonant triad with two conservation laws coined as energy and enstrophy \cite{Pedlosky1987,Reznik1993,Lynch2003}. It is well-known that the triad equations have a Nambu or Lie-Poisson structure \cite{Holm2008geometric} with the same bracket as in rigid body dynamics. In the atmosphere Rossby waves are forced by constant orographic inhomogeneities \cite{Lynch2003, Lynch2009} and the amplitudes show a typical recharge-discharge cycle. 


The aim of  this paper is to present Nambu representations for forcings in the real wave triad equations.  
Harris et al. \cite{HARRIS20124988} have determined stability and boundedness properties of the equation in complex form with a forcing applied to the unstable mode. The Hamiltonian for our system is the unforced energy plus functions of the unforced amplitudes. For the second conservation laws, the Casimir functions in Hamiltonian theory, we replace enstrophy by relations obtained by the unforced equations. The Nambu bracket is the same as in the unforced equations. In simulations with an intermediate wavenumber forcing, a typical recharge process is induced. Recharge cycles are common in geophysical fluid dynamics and typically modeled as nonlinear oscillators (see the models for baroclinic storms \cite{Ambaum2014}, convection \cite{Yano2012}, and wave-mean flow interaction \cite{Blender2013}). To demonstrate the usefulness of the Nambu representation we derive the corresponding equations by approximating the conservation laws for the recharge cycle.



The paper is organized as follows: In Section \ref{sec_unforced_triads} the geometric representation of spherical Rossby wave triads without forcing is revisited. In Section \ref{sec_forc_triads} triads with different forcings are described as a Nambu systems. For the  intermediate wavenumber forcing a recharge cycle is obtained and approximated as a canonical Hamiltonian system. In Section \ref{sec_SumDis} the results are summarized and discussed. 

\section{Spherical Rossby wave triads without forcing} \label{sec_unforced_triads}

Large scale atmospheric dynamics is governed by the barotropic vorticity equation. For small amplitudes linear solutions are given by noninteracting Rossby-Haurwitz waves. 
A triad of these waves is given when they satisfy resonance conditions \cite{Pedlosky1987,Reznik1993,Lynch2003}.  The three waves are decoupled from the rest and energy is only exchanged within this triad. Note that the interaction within a triad is weakly nonlinear and only valid for moderate amplitudes. For higher amplitudes the decoupling breaks down, the waves interact with all others, and the flow becomes turbulent.  


Reznik et al. \cite{Reznik1993} derived the amplitude equations of spherical Rossby wave triads by a multiple time scale analysis of the barotropic vorticity equation (BVEQ) (see also \cite{Pedlosky1987}, \cite{Bender1999}). The equations are real and the phases are disregarded. The amplitudes of the waves in a triad vary slowly compared to the wave frequency. The nonlinearity in the BVQE requires that three waves form a triad if the meridional wave numbers $m$ and the frequencies $\omega$ satisfy the resonance conditions $m_1+ m_2 = m_3$, and $\omega_1 + \omega_2 = \omega_3$, where $\omega_1= \omega_1(n_1,m_1)$, etc.,  with the total wave number $n$ and and the linear dispersion relation $\omega(n,m)$ of the Rossby waves. 

The  amplitude equations are not determined in the BVEQ directly, but by the condition in the expansion which requires that the perturbations remain bounded for long times. This leads to the three equations for the slow amplitudes $A_1$, $A_2$, and $A_3$ in a triad
\begin{equation} \label{AAAZN}
\eqalign{
N_1 \frac{\rmd A_1}{\rmd T} &= Z(N_2-N_3)A_2 A_3 \\
N_2 \frac{\rmd A_2}{\rmd T} &= Z(N_3-N_1)A_1 A_3 \\
N_3 \frac{\rmd A_3}{\rmd T} &= Z(N_1-N_2)A_1 A_2,
}
\end{equation}
The parameters $N_i=n_i(n_i+1)$ are determined by the total wave numbers of the Rossby waves and $Z$ is the interaction coefficient \cite{Reznik1993}. Note that the phase space divergence of the equations (\ref{AAAZN}) for the vector $\mathbf{A}=(A_1,A_2,A_3)$ vanishes, $\nabla \cdot \dot{\mathbf{A}} =0$.

\subsection*{Nambu  representation}

The system (\ref{AAAZN}) has two conservation laws, the energy
\begin{equation} \label{HA123}
	 H = \frac{1}{2}(N_1 A_1^2 + N_2 A_2^2 + N_3 A_3^2) \\
\end{equation}
and the enstrophy 
\begin{equation}   \label{CA123}
 	C = \frac{1}{2} (N_1^2 A_1^2 + N_2^2 A_2^2 + N_3^2 A_3^2).
\end{equation}
Due to the conservation laws the equations are integrable. Exact solution are given in terms of Jacobian elliptic functions. 

The amplitude equations can be formulated as a Nambu system for the state space vector  $\mathbf{A}$
\begin{equation} \label{dAdtNambu}
	\frac{\rmd \mathbf A}{\rmd t} = \frac{Z}{N_1 N_2 N_3} \nabla C \times \nabla H
\end{equation}
where the $\nabla$-operator represents $A$-derivatives. 
The dynamics of an arbitrary function $F(A_1,A_2,A_3)$  is given in terms of a Nambu bracket
\begin{equation} \label{FgCHf}
	\frac{\rmd F}{\rmd t} = \{F,C,H\} 
\end{equation}
which is the rigid body Nambu bracket up to a constant factor  
\begin{equation}  \label{FABzn}
	\{F,A,B\} =  \frac{Z}{N_1 N_2 N_3} \nabla F \cdot \nabla A \times \nabla B
\end{equation}
%
%
A Nambu representation is suggested in \cite{Holm2008geometric} and interpreted geometrically by \cite{Holm2002stepwise}.  A Lie-Poisson structure is obtained by $\{F,H\}_C = \{F,C,H\}$ \cite{Takhtajan1994}, where $C$ is a Casimir.  
For the geometric visualization of phase space dynamics it is helpful that the equations are unchanged for linear combinations of the conservation laws, e.g. $\nabla C \times \nabla(H+C)$. 

\subsection*{Standard amplitudes}

It is convenient to transform the dynamic equations to standard amplitudes (see e.g. \cite{Lynch2009}). Here we consider the wave number ordering $N_1 < N_2 <N_3$. 
\begin{equation} \label{YQX}
	Y_n= \sqrt{  \frac{N_n}{D_n} }  \tilde{Q}  A_n, 
	\qquad n=1, 2, 3
\end{equation} 
where
\begin{equation} \label{YQX}
	\tilde{Q} = \sqrt{\frac{D_1 D_2 D_3}{N_1 N_2 N_3} }
\end{equation} 
and
\begin{equation} \label{DDD}
	D_1=N_3-N_2, \quad D_2=N_3-N_1, \quad D_3=N_2-N_1
\end{equation} 
which are all positive. 
The dynamical equations for the standard variables are 
\begin{equation} \label{YYY}
\eqalign{
 \frac{\rmd Y_1}{\rmd t} &= -Y_2 Y_3 \\
 \frac{\rmd Y_2}{\rmd t} &=  Y_1 Y_3 \\
 \frac{\rmd Y_3}{\rmd t} &= -Y_1 Y_2
}
\end{equation}
The unstable mode is the intermediate wavenumber amplitude $Y_2$.  

The conservation laws  (\ref{HA123}, \ref{CA123}) for the standard variables  are the Hamiltonian
\begin{equation} \label{HY123}
	 H = \frac{1}{2}(D_1 Y_1^2 + D_2 Y_2^2 + D_3 Y_3^2) \\
\end{equation}
and the Casimir
\begin{equation}   \label{CY123}
	C = \frac{1}{2} (N_1 D_1 Y_1^2 + N_2 D_2 Y_2^2 + N_3 D_3  Y_3^2) \\
\end{equation}
For the state vector  $\mathbf{Y}=(Y_1,Y_2,Y_3)$ the Nambu form (\ref{dAdtNambu}) reads as (the interaction coefficient $Z$ is omitted in the following, since it can be included in the time scale)
\begin{equation} \label{dYdtNambu}
	\frac{\rmd \mathbf Y}{\rmd t} = \frac{1}{\tilde{Q}} \nabla C \times \nabla H
\end{equation}
with the $\nabla$-operator representing $Y$-derivatives. 

In the analysis below where we consider forced equations we will use Casimir functions based on conservation laws derived in the unforced equations. The advantage of these functions is that they can be derived in the pair of the unforced equations.  A well-known conservation law of (\ref{YYY}), based on the Manley-Rowe relations, is 
\begin{equation} \label{J2}
	J^2_2 = Y_3^2-Y_1^2
\end{equation}
This can be derived by integrating $Y_1 dY_1=Y_3 dY_3$ in the equations for $Y_1,Y_3$.
This conservation law allows an alternative Nambu form of the conservative equations 
\begin{equation} \label{dYdtNambuJH}
	\frac{\rmd \mathbf Y}{\rmd t} =  \nabla C_2 \times \nabla H
\end{equation}
where
\begin{equation} \label{C2}
	C_2 = \frac{1}{2 D_2} (Y_3^2-Y_1^2)
\end{equation}
is a Casimir function. Here we incorporasted the factor $1/\tilde{Q}$ in (\ref{dYdtNambu}) in the Casimir. The bracket notation for (\ref{dYdtNambuJH}) is 
\begin{equation} \label{FC2H}
	\frac{\rmd F}{\rmd t} = \{F,C_2,H\}
\end{equation}
with the rigid body Nambu bracket for an arbitrary function $F(Y_1,Y_2,Y_3)$ 
\begin{equation}  \label{FAB}
	\{F,A,B\} =  \nabla F \cdot \nabla A \times \nabla B 
\end{equation}
%
%
\section{Forced triad} \label{sec_forc_triads}

We consider constant forcings in the three equations separately 
\begin{equation} \label{YYYforce123}
\eqalign{
 \frac{\rmd Y_1}{\rmd t} &= -Y_2 Y_3 + f_1\\
 \frac{\rmd Y_2}{\rmd t} &=  Y_1 Y_3 + f_2\\
 \frac{\rmd Y_3}{\rmd t} &= -Y_1 Y_2 + f_3
}
\end{equation}
Since we did not include friction, phase space divergence $\nabla \cdot \dot{\mathbf{Y}}$ vanishes. To derive the geometric representation we consider the three forcing terms separately.


\subsection{Forcing of the amplitude $Y_1$}
%
First we restrict the forcing to the small wavenumber amplitude $Y_1$  
\begin{equation} \label{Y1force}
\eqalign{
 \frac{\rmd Y_1}{\rmd t} &= -Y_2 Y_3 + f_1 \\
}
\end{equation}
while the forcings in $Y_2$ and $Y_3$ are disregarded, $f_2=f_3 = 0$ in (\ref{YYYforce123}).
The conservation law derived in the unforced $Y_2$ and the $Y_3$ equations is 
\begin{equation} \label{J1}
	J_1^2 = Y_2^2+Y_3^2
\end{equation}
This is used to define  the Casimir
\begin{equation} \label{C1}
	C_1 = 1/(2 D_1) (Y_2^2+Y_3^2)
\end{equation}
The 'forced Hamiltonian' is 
\begin{equation} \label{Hf1}
	 H^f_1= H - f_1 D_1 \arcsin (Y_2/J_1)
\end{equation} 
where $H$ is (\ref{HY123}). The Nambu representation for the system with a forcing in $Y_1$ only is
\begin{equation} \label{YC1Hf1}
	\frac{\rmd \mathbf{Y}}{\rmd t} = \nabla C_1 \times \nabla H^f_1  
\end{equation}
For an arbitrary function $F(Y_1,Y_2,Y_3)$ the dynamics is $\rmd F/\rmd t = \{F,C_1,H^f_1\}$ 
with the rigid body bracket (\ref{FAB}).

\subsection{Forcing of the amplitude $Y_2$}
%
Here we consider a constant forcing $f_2$ in the unstable mode $Y_2$ while $f_1=f_3=0$
%
\begin{equation} \label{YYYforce2}
\eqalign{
 \frac{\rmd Y_2}{\rmd t} &=  Y_1 Y_3  + f_2\\
}
\end{equation}
For the Nambu representation we use the Casimir (\ref{C2}) and the 'forced Hamiltonian'. 
\begin{equation} \label{Hf2}
	 H^f_2= H+ f_2 (D_2/2) \ln (Y_1+Y_3)^2
\end{equation} 
%
The Nambu representation for the system with a forcing in $Y_2$ only is
\begin{equation} \label{YC1Hf1}
	\frac{\rmd \mathbf{Y}}{\rmd t} = \nabla C_2 \times \nabla H^f_2  
\end{equation}
and for am arbitrary function $F$ the bracket for the forcing $f_2$ is 
$\rmd F/\rmd t = \{F,C_2,H^f_2\}$. 

\begin{figure}
\includegraphics[width=0.8\textwidth]{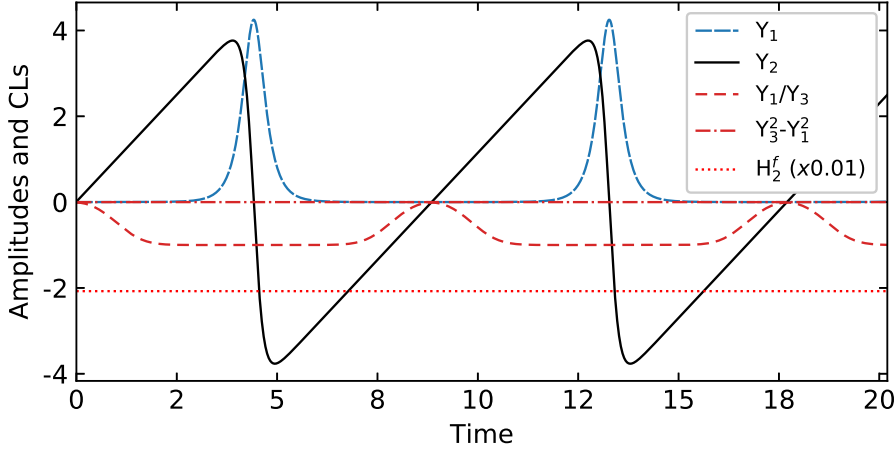} 
\caption{Triad with the forced amplitude $Y_2$. 	
Initial conditions are $Y_1=Y_2=10^{-5},Y_3=-10^{-3}$. 
Amplitudes: $Y_1$ (dashed blue), $Y_2$ (solid black), and ratio $r=Y_1/Y_3$ (dashed red). 
Conservation laws:   $Y_3^2-Y_1^2$ in (\ref{J2}) (dash-dotted red) and 
forced energy $H^f_2$ in (\ref{Hf2})  (dotted red, reduced by $0.01$).
}
\label{fig:forced_Y} 
\end{figure}

The conservation laws  $J^2_2 = Y_3^2-Y_1^2$ (\ref{J2}) and (\ref{Hf2}) allow wide  excursions of $Y_1$ and $Y_3$ with opposite sign. Therefore, the dynamics differs from the $f_1$-forcing  where $J_1^2 = Y_2^2+Y_3^2 $ (\ref{J1}) remains constant. This leads to  a recharge behavior  for the $f_2$-forcing (Fig.~\ref{fig:forced_Y}) which is considered in more detail below. 


\subsection{Forcing of the amplitude $Y_3$}

If the forcing is applied in $Y_3$ only
\begin{equation} \label{Y3force}
	\eqalign{
	 \frac{\rmd Y_3}{\rmd t} &= -Y_1 Y_2 + f_3
	}
\end{equation}
the Casimir used in the Nambu representation is derived in the unforced $Y_1, Y_2$ equations
\begin{equation} \label{C3}
	C_3 = 1/(2 D_3) (Y_1^2+Y_2^2)
\end{equation}
The forced Hamiltonian is 
\begin{equation} \label{Hf3}
	 H^f_3 = H + f_3 D_3 \arcsin (Y_2/J_3)
\end{equation} 
with 
\begin{equation} \label{J3}
	J_3^2 = Y_1^2+Y_2^2
\end{equation}
%
The opposite signs in (\ref{Hf1}) and (\ref{Hf3}) originate in the definitions (\ref{DDD}) with $D_3+D_1=D_2$.
The bracket for the forcing $f_3$ is $\rmd F/\rmd t = \{F,C_3,H^f_3\}$. 
%

%
%

\subsection{Recharge process in the forced $Y_2$-equation}
 
The recharge process for a forcing of the intermediate wavenumber in the triad  is demonstrated in a numeric simulation of (\ref{YYYforce2}) with the wavenumbers $n_1=6, n_2=7, n_3=8$. This reveals a recharge cycle with a gradual increase of the forced amplitude, a sudden burst in the unforced waves  (denoted as perturbations here),  a reversal of the forced wave and a subsequent recovery (see Fig.~\ref{fig:forced_Y},  compare also Figure 4 in \cite{Lynch2009} with a forcing in the complex equations). Harris et al.  \cite{HARRIS20124988} underline, that this forcing is not a source of energy. 


The process is characterized by an opposite sign of the perturbations, $Y_1 \approx -Y_3$ (the ratio $r=Y_1/Y_3$ in Fig.~\ref{fig:forced_Y} tends to $-1$ during the recharge). 
When this relation holds, the perturbations grow  with a rate proportional to $Y_2$ according to  
$\rmd Y_1/\rmd t = -Y_2 Y_3 $,
while $Y_2$ is reduced by a positive value of $-Y_1 Y_3$ (\ref{YYYforce2}). 
To describe this process we approximate the equations for a small deviation $\delta$ from the opposite sign in the perturbations, $Y_3 = -Y_1 + \delta$.  The variables $(Y_1,Y_2,Y_3)$ in (\ref{YYYforce2}) are replaced by $(Y_1,Y_2,\delta)$ with the equations 
\begin{equation} \label{YYdf}
\eqalign{
 \frac{\rmd Y_1}{\rmd t} &= Y_1 Y_2 - Y_2 \delta \\
 \frac{\rmd Y_2}{\rmd t} &=  -Y_1^2 +  Y_1 \delta   + f\\
 \frac{\rmd \delta}{\rmd t} &= -Y_2 \delta
}
\end{equation}
During the recharge interval when $Y_2>0$, the sum $\delta = Y_1+Y_3$ decays (\ref{YYdf}) and the perturbations align to $Y_1 = - Y_3$. 

These equations can be obtained in a Nambu form if we approximate the two conservation laws (\ref{C2}, \ref{Hf2})  in the same Nambu operator (\ref{FAB}). We approximate the two conservation laws to order $O(\delta)$. The conservation law corresponding to the Casimir is
\begin{equation} \label{CJd}
	C_2 = - \frac{1}{D_2} Y_1 \delta
\end{equation}
and the Hamiltonian is
\begin{equation} \label{HYfd}
	 H^f_2 = H+ f (D_2/2) \ln \delta^2
\end{equation} 
The main equations governing the recharge process are obtained if we ignore $\delta$ as a degree of freedom in  (\ref{YYdf}) 
\begin{equation} \label{YYf}
	\eqalign{
 	\frac{\rmd Y_1}{\rmd t} &= Y_1 Y_2  \\
 	\frac{\rmd Y_2}{\rmd t} &=  -Y_1^2  + f
	}
\end{equation}

Similar nonlinear oscillators have been suggested by  \cite{Ambaum2014} and \cite{Yano2012}.
This reduced system possesses a canonical Hamiltonian representation in terms of the variables $Y_2$ and $\eta=(1/2) \log Y_1^2/2 $, which describes the perturbation intensity. We can write it as a symplectic system
\begin{equation}
\left(
	 \begin{array}{c}
	\rmd Y_2/\rmd t  \\
	\rmd \eta/\rmd t 
	\end{array}
\right )  =
\left ( \begin{array}{ccc}
0 & -1 \\
1 & 0
\end{array} 
\right )
\left(
	 \begin{array}{c}
	\partial /\partial Y_2 \\
	\partial /  \partial \eta
	\end{array}
\right )
H_r
\end{equation}
with the Hamiltonian $H_r = (1/2) Y_2^2 + \exp(2 \eta) - f \eta$.

\section{Summary and Discussion} \label{sec_SumDis}

In this paper we derived Nambu representations for constant forcings in the three wave equations for real amplitudes. A geophysical example are the amplitudes equations for resonant Rossby wave triads \cite{Pedlosky1987, Reznik1993}. Without forcing these equations possess two conservation laws, coined as energy and enstrophy. The  dynamics can be written in a Nambu form with the canonical Nambu bracket (this is already known from  \cite{Holm2008geometric}), thus the triads are mathematically equivalent to the rigid body dynamics. An alternative Nambu description is given if enstrophy is replaced by a geometric conservation law based on the Manley-Rowe relations. 

For forcings in the three  equations Nambu forms are obtained with the Hamiltonian extended by perturbations given by functions of the unforced amplitudes. The second conservation laws are based on relations obtained in the unforced equations. The forcing of the intermediate (unstable) wavenumber is considered in detail since these equations yield a recharge process. This is characterized by an opposite alignment of the unforced amplitudes. The approximated equations are obtained in a Nambu representation with expanded conservation laws. 

The main result is that we could describe a constantly forced system in a Nambu representation. Note that we did not include dissipation associated with phase space convergence  which needs to be  included as a separate gradient term \cite{kaufman84}. A representation of a physical system in terms of its conservation laws in a Nambu form is useful for the following reasons: (i) Time evolution is interpreted as a nondivergent flow in phase space and conservation laws act as stream-functions, (ii) Consistent approximations are  obtained by approximating the conservation laws \cite{NevirSommer2009}, (iii) Conservative numeric codes can be derived by symmetry properties of the Nambu bracket \cite{Salmon2005}. Further applications of conservation laws are in nonlinear stability by the Energy-Casimir method \cite{blumen1968stability}, and statistical mechanics  \cite{Bouchet2012}. 
For a brief review on applications of Nambu mechanics in geophysical fluid dynamics see the corresponding chapter in \cite{Lucarini2014mathematical}.

As an outlook this finding gives support to a modeling strategy which is purely based on conservation laws. Blender and Badin \cite{BlenderBadin2015} have demonstrated that the Rayleigh-B\'{e}nard equations can be derived based on a bilinear structure of a conservation law (the Casimir) in the canonical Nambu bracket. 
Kaltsas and Throumoulopoulos \cite{Kaltsas2019} could derive new conservative equations in magneto-dynamics based on this idea. Very promising, but less pursued in hydrodynamics, is the parameterization of processes where we know exact conservation laws (see  \cite{frank2011unifying} for chemical reactions).

\section*{Acknowledgements}
The study was partly funded by the Deutsche Forschungsgemeinschaft (DFG, German Research Foundation) under Germany‘s Excellence Strategy – EXC 2037 'CLICCS - Climate, Climatic Change, and Society' – Project Number: 390683824, contribution to the Center for Earth System Research and Sustainability (CEN) of Universit\"at Hamburg. RB ackowledges support by the German Reserch Foundation (DFG, Grant BL 516/3-1). 

\section*{References}
\bibliographystyle{iopart-num}
\bibliography{Blender-Triad-Nambu}

\end{document}